\documentclass{aa}
\usepackage[utf8]{inputenc}

\usepackage{graphicx}
\usepackage{txfonts}
\begin{document} 

   \title{Amplitude variations of modulated RV Tauri stars support the dust obscuration model of the RVb phenomenon}
   \titlerunning{Amplitude variations of RVb stars}
   
   \authorrunning{L.L. Kiss \& A. B\'odi}

   \author{L.L. Kiss
          \inst{1,2} \and
          A. B\'odi
          \inst{1,3}
          }

   \institute{
   Konkoly Observatory, Research Centre for Astronomy and Earth Sciences, Hungarian Academy of Sciences, H-1121 Budapest, Konkoly Thege M. \'ut 15-17, Hungary\\
   \email{kiss.laszlo@csfk.mta.hu}
         \and
   Sydney Institute for Astronomy, School of Physics A28, University of Sydney, NSW 2006, Australia
         \and
   Department of Experimental Physics and Astronomical Observatory, University of Szeged, H-6720 Szeged, D\'om t\'er 9., Hungary
             }

\date{July 2017}

\abstract{
    RV~Tauri-type variables are pulsating post-AGB stars that evolve rapidly through the instability strip after leaving the Asymptotic Giant Branch. Their light variability is dominated by radial pulsations. Members of the RVb subclass show an additional variability in form of a long-term modulation of the mean brightness, for which the most popular theories all assume binarity and some kind of circumstellar dust. Here we address if the amplitude modulations are consistent with the dust obscuration model.
}{
    We measure and interpret the overall changes of the mean amplitude of the pulsations along the RVb variability.
}{
    We compiled long-term photometric data for RVb-type stars, including visual observations of the American Association of Variable Star Observers, ground-based CCD photometry from the OGLE and ASAS projects and ultra-precise space photometry of one star, DF~Cygni, from the { \it Kepler} space telescope. After converting all the observations to flux units, we measure the cycle-to-cycle variations of the pulsation amplitude and correlate them to the actual mean fluxes. 
}{
    We find a surprisingly uniform correlation between the pulsation amplitude and the mean flux: they scale linearly with each other for a wide range of fluxes and amplitudes. It means that the pulsation amplitude actually remains constant when measured relative to the system flux level.
    The apparent amplitude decrease in the faint states has long been noted in the literature but it was always claimed to be difficult to explain with the actual models of the RVb phenomenon. Here we show that when fluxes are used instead of magnitudes, the amplitude attenuation is naturally explained by periodic obscuration from a large opaque screen, one most likely corresponding to a circumbinary dusty disk that surrounds the whole system. 
}{}

   \maketitle

\section{Introduction}

The RV~Tauri variables constitute a small group of pulsating stars with some dozen known members in the Milky Way and a similar number of variables in the Magellanic Clouds. They are F, G and K-type supergiants that form the high-luminosity extension of Population II Cepheids in the classical instability strip (Wallerstein 2002). The rarity of RV~Tau-type variables can be explained by their evolutionary state given that these objects are post-AGB stars rapidly evolving in the Hertzprung-Russell diagram blueward from the Asymptotic Giant Branch (AGB) on the timescales of 10$^3$--10$^4$ yrs (Bl\"ocker 1995). During this evolution, post-AGB stars cross the instability strip, where they become pulsationally unstable (Fokin 1994, Fokin et al. 2001, Aikawa 2010). The most regular, though far from being  Cepheid-like, high-amplitude light curves belong to the RV~Tauri stars, which are located in cooler part of the post-AGB instability strip, between 5000 K and 6000 K (Kiss et al. 2007, B\'odi et al. 2016).

The most distinct feature of the RV~Tau-type variables is the presence of alternating minima of the pulsations (meaning that every second light curve minimum is shallower), with typical (double) periods from 30 days to 90 days. The periodicity is not strict, as the cycle-to-cycle variations can be quite significant, and in some cases, it has been shown to originate from low-dimensional chaos (e.g. Buchler et al. 1996). In addition to the pulsations, some RV~Tau stars show long-term modulation of the mean brightness, with periods of 700-2500 days. The absence or presence of the slow modulation is the basis for classifying the stars into the RVa and RVb photometric subclasses, respectively.

The RVb phenomenon has been connected to the fact that these stars are typically redder than the RVa-type variables and the role of circumstellar dust shells was presumed (Lloyd Evans 1985). More recent studies interpret the RVb phenomenon with periodic obscuration events in binary systems (Fokin 1994, Pollard et al. 1996, Van Winckel et al. 1999, Fokin et al. 2001, Maas et al. 2002, Gezer et al. 2015). In that picture all RVb stars are binaries, surrounded by a large opaque screen, relative to which the pulsating component changes its position during the orbit. The presence of a dusty disk is indeed seen in the infrared part of the Spectral Energy Distribution (see Gezer et al. 2015 for the latest analysis that used WISE data and references therein). In addition to the disk, it has also been speculated that interactions of the components may lead to changes in the pulsation amplitude (Pollard et al. 1996, Maas et al. 2002, Pollard et al. 2006), which could further complicate the models of these stars.

The fact that the pulsations in the light curve appear to visually decrease in amplitude in the faint states of some RVb variables has been regularly noted in the literature. Fokin (1994) compiled data for a dozen stars and noted that all showed systematically lower primary light amplitudes in the minimum of the secondary variation. Pollard et al. (1996) discussed in details the well-expressed pulsation amplitude decrease in the faint states of U~Mon and AI~Sco (i.e. around the RVb-type minima), which they assumed to be caused by some sort of dynamical interaction between the pulsating and the companion stars, which affects both the pulsations and the mass-loss processes. Pollard et al. (1996) explicitly argued that pure obscuration by dust should not decrease the amplitude of the pulsations. This line of argumentation was later adopted by Van Winckel et al. (1999) and Maas et al. (2002), who claimed that the detected amplitude change is difficult to account for in a simple geometric picture. Later, Pollard et al. (2006) reiterated the main conclusion that enhanced mass-loss or binary interaction would be needed around the periastron of U~Mon to explain the apparent amplitude damping. Most recently, Percy (2015) investigated the amplitude variability of 42 RV~Tau variables and found that the pulsation amplitude can change by factors of up to 10, on median time-scales of about 22 pulsation periods. He concluded that the cause of the pulsation amplitude variations remains unknown. 
     
 Despite the efforts in the past decades,  high-quality photometry of RV~Tau-type stars is still very rare. In the original {\it Kepler} field, there is only one star, DF~Cygni, a moderately bright RVb-type variable, for which independent analyses were recently published by B\'odi et al. (2016) and Vega et al. (2017). Both studies pointed out that DF~Cygni shows very rich behaviour on all timescales. While B\'odi et al. (2016) put their emphasis on detecting evidence of strong non-linear effects that are directly observable in the { \it Kepler} light curve, Vega et al. (2017) used {\it Kepler} data to argue for binarity as the main cause of long-period variability in DF~Cyg. The latter authors also noted that when the light curve is measured in fluxes, the reduction of the pulsation amplitude in the faint state of DF~Cyg is exactly the same $\sim$90\% than the overall fading by 90\% during the RVb minimum. The exact correspondence of the two can be naturally explained by 90\% obscuration of the pulsating stellar disk by a very large opaque screen during every orbit, so that the local changes around the mean (i.e. the short-period pulsations) are in fact constant when considered relative to the mean brightness of the system.  
 
 This idea is the very inspiration of the present paper. We believe the earlier studies overlooked a very simple and yet important point by sticking to the inverse logarithmic magnitude system in the variable star analyses, rather than using the physically meaningful flux units, which scale linearly with the photon counts. As we will demonstrate for the best observed RVb-type variables, a very convincing and ubiquitous correlation exists between the flux amplitudes and the mean flux levels for all RVb stars, which indicates that dust obscuration, most likely by circumbinary disks, can indeed be the universal explanation for the RVb phenomenon. 
 
\section{Data and methods}

Given the rarity of the RV~Tau-type stars, there are not many variables with extensive observations. We have surveyed the The International Variable Star Index (VSX database\footnote{http://www.aavso.org/vsx/}) and the literature for well observed and characterized targets. We ended up with three major sources of RVb photomeric data. First, we checked the database of visual observations of the American Association of Variable Star Observers (The AAVSO International Database). In total, we found eight stars with several cycles of the RVb variability and duty cycle in excess of 75\%, which was found to be necessary to measure the amplitudes of individual pulsation cycles. Then we checked the online catalogue of the All Sky Automated Survey (ASAS, Pojmanski 2002) and found $V$-band data for three stars. Finally, we surveyed the database of the Optical Gravitational Lensing Experiment (OGLE) project, where the OGLE-III Catalog of Variable Stars contains Type II Cepheids (including RV~Tau-type variables) in the Large Magellanic Cloud (Soszynski et al. 2008), Small Magellanic Cloud (Soszynski et al. 2010) and the Galactic Bulge (Soszynski et al. 2011, 2013). Here we found useful $I$-band data for six stars in the Bulge and one star in the LMC, all catalogized as `RVb' variables by the OGLE team. 

\begin{table}[t]
\caption{The studied sample of RVb stars. T$_{\rm obs}$ and N$_{\rm obs}$ are the time-span and the total number of observations. P$_{\rm pul}$ and P$_{\rm mod}$ refer to the periods of pulsation and RVb-type modulation, respectively, as measured from the analysed data. The OGLE variable names were shortened by omitting ``OGLE-'' in front of the shown identifiers.}
\label{table:stars}
\centering
\setlength{\tabcolsep}{3pt}
\begin{tabular}{lrrrrr}
\hline\hline
Name & T$_{\rm obs}$ & N$_{\rm obs}$ & P$_{\rm pul}$ & P$_{\rm mod}$ & Source \\
   & (d) &  & (d) & (d)  & \\
\hline
IW~Car  & 18120 & 4685 & 71.98 & 1449 & AAVSO\\
        & 3300 & 2179 & 72.2 & 1470 & ASAS\\
SX~Cen  & 22409 & 1320 & 32.88 & 602 & AAVSO\\
        & 3296 & 1153 & 33.01 & 610 & ASAS\\
DF~Cyg  & 17074 & 5924 & 49.82 & 780 & AAVSO\\
        & 1470 & 66533 & 49.84 & 786 & { \it Kepler}\\
SU~Gem  & 14783 & 2228 & 49.92 & 682 & AAVSO\\
U~Mon   & 46283 & 48019 & 91.48 & 2451 & AAVSO\\
AR~Pup  & 14998 & 1450 & 76.66 & 1194 & AAVSO \\
        & 3299 & 1086 & 76.34 & 1178 & ASAS\\
AI~Sco & 19538 & 1408 & 71.64 & 977 & AAVSO\\
RV~Tau  & 40020 & 14976 & 78.48 & 1210 & AAVSO\\
BLG-T2CEP-177 & 2836 & 742 & 92.44 & 2970 & OGLE\\
BLG-T2CEP-215 & 2829 & 814 & 55.74 & 958 & OGLE\\
BLG-T2CEP-345 & 2830 & 1344 & 73.64 & 1100 & OGLE\\
BLG-T2CEP-350 & 2776 & 1026 & 87.20 & 722 & OGLE\\
BLG-T2CEP-352 & 4404 & 978 & 103.78 & 543 & OGLE\\
BLG-T2CEP-354 & 3232 & 533 & 66.46 & 951 & OGLE\\
LMC-T2CEP-200 & 4494 & 917 & 69.86 & 850 & OGLE\\
\hline
\end{tabular}
\end{table}

The final sample contains 19 datasets for 15 RVb stars, which is - to our knowledge - the most extensive collection of this kind in the literature. In Table~\ref{table:stars} we list the main characteristics of the data: the total time-span of the individual light curves, the number of measurements per star and two period values: one for the pulsations and one for the RVb modulation. The listed values were measured with standard Fourier analysis, using Period04 of Lenz \& Breger (2005). Because of the alternating minima, the highest peak in the Fourier spectra always corresponded to the half-period of the RV~Tau cycle, hence we doubled the adopted period of pulsations. The only exception is IW~Car, whose light curve does not show any alternation in the minima, hence we adopted the single cycle period for its pulsations. The typical uncertainties in the quoted values are in the order of the last digit shown in Table~\ref{table:stars}. Note that in principle, the pulsation period could be determined to more decimals, however, the seemingly irregular variations in the period/phase (B\'odi et al. 2016) would pose strong limitations to all future phase predictions. Our measured values are all in good agreement with previous period determinations in the literature (e.g. Pollard et al. 1996, Kiss et al. 2007, Percy et al. 2015).

\begin{figure*}
    \includegraphics[width=18cm]{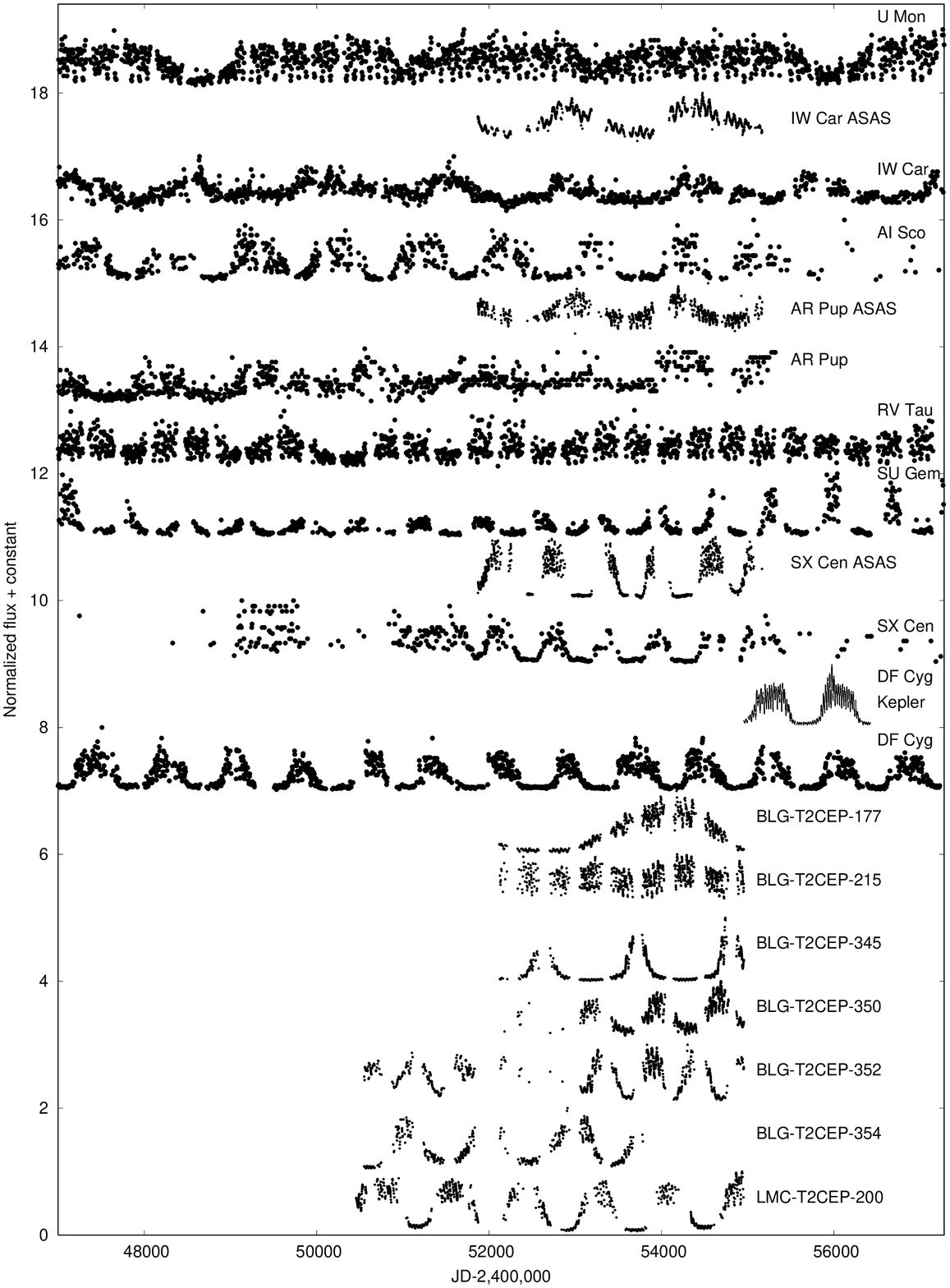}
      \caption{A gallery of RVb light curves, converted to fluxes and then normalized to unity.}
      \label{aavso}
   \end{figure*}

The visual data have first been binned to average out the errors of the individual observations (typically $\pm$0.2-0.3 mag per point). Because of the short pulsation periods, the bin sizes were selected between 3 days and 5 days, depending the pulsation period. This way we avoided strong phase smearing due to the binning, that could have led to undersampled light curves of pulsations. The CCD observations were only checked for outliers, identified by visual inspection of the data. Both ASAS and OGLE have daily cadence and no binning was applied to those data.

\begin{figure}
   \centering
   \includegraphics[width=9cm]{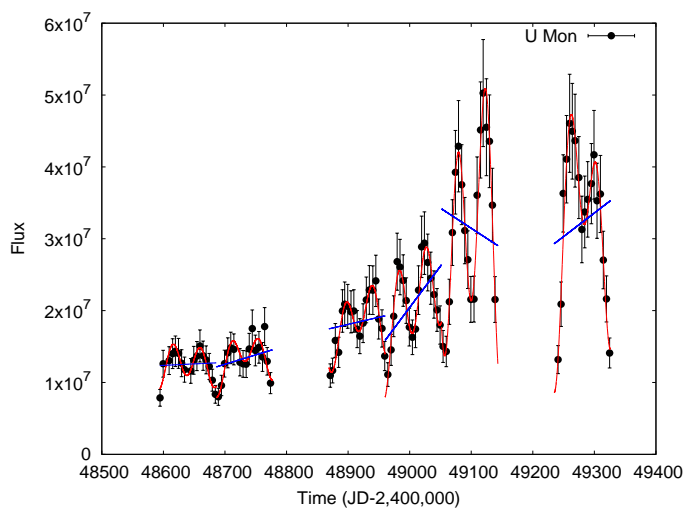}
      \caption{A short subset of the AAVSO light curve of U~Mon (black dots), overplotted with the fitted linear term (blue lines) and the sum of two sine waves (red dotted lines). The damping of the pulsation amplitude during the minimum phase (JD 48,600-48,800) of the RVb variation clearly visible.}
         \label{umon}
   \end{figure}

Following the idea outlined by Vega et al. (2017), we converted every light curve to flux units with the conversion formula $f=10^{-0.4\times({\rm magn.}-25)}$, adapting 25.0 as the arbitrary zero-point. To demonstrate the overall continuity and typical duty cycle of the data, in Fig.~\ref{aavso} we show a gallery of the full sample in normalized flux units (unit flux corresponds to the global maximum brightness for each star). While we plotted the entire ASAS, { \it Kepler} and OGLE light curves, the AAVSO data are only shown partially, in 10,000-days long subsets. For the overwhelming majority of stars the amplitude variability of the short-period pulsations is very much apparent, with the highest amplitudes always appearing in the bright phases of the RVb variability. Exceptions are IW~Car, AR~Pup (in the ASAS data) and OGLE-BLG-T2CEP-215, which seem to have more stable pulsations with RVb modulations in the same range as the pulsation amplitudes.

We proceeded in the analysis by measuring the local amplitudes of each pulsation cycle. For this, each dataset was split into subsets that covered exactly one pulsation cycle (two consecutive maxima and minima of the alternating depths) and then we fitted the following function to each of the subsets:

\begin{equation}
    f(t-t_0)=f_0+k (t-t_0) + \sum_{n=1}^{2}A_n \sin  \Bigg ( \frac{2\pi (t-t_0)}{n P_{\rm pul}}+\varphi_n \Bigg )
\end{equation}

\noindent where $t$ and $t_0$ are time and the first time epoch of the subset, respectively, $f_0$, $k$, $A_n$ and $\varphi_n$ ($n=1,2$) are the fitted parameters (for IW~Car, $n=1$ was used). The zero-point and the linear term were introduced to include the local variation of the slow RVb modulation, while the two-component Fourier-polynomial is the approximation of the light curve shape of the pulsations. The latter is admittedly a simplified description of the light curve, however, here we are more interested in the global average changes of the pulsation amplitudes than in the perfect fits of the individual cycles. We have also experimented with higher-order polynomials, but none of the results changed significantly, hence we restricted the analysis to the simplest form. An illustration of the fitting procedure is shown in Fig.~\ref{umon}, where the red curve shows the individual fits of Eq.\ 1 throughout a whole ascending branch of the RVb cycle. The blue line separately indicates the linear term. The main conclusion here is that the two-component harmonic fit describes the observations very well.     

After fitting Eq.\ 1 to a given subset, we subtracted the linear term and computed the two extrema of the residual polynomial (except for the {\it Kepler} data of DF~Cyg, where the extrema were determined from the residual of the observational points). Their difference was taken as the full pulsation amplitude in the given subset, while the median of the subset was chosen to represent the average flux of the star. The relationship that was found between these amplitudes and mean fluxes and its properties represent the main results of this paper. 

\section{Results}

    \begin{figure}
   \centering
    \includegraphics[width=9cm]{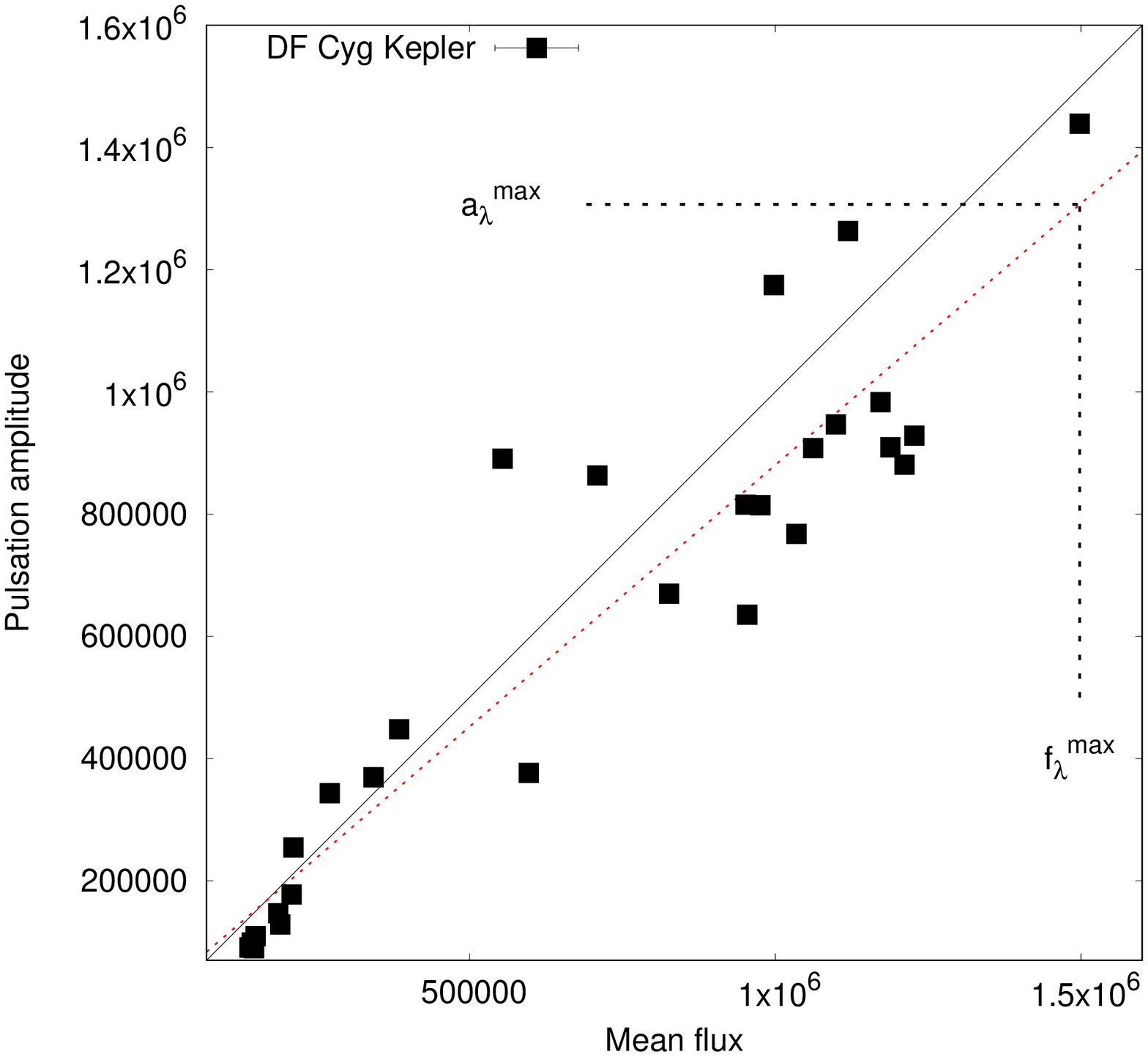}
    \includegraphics[width=9cm]{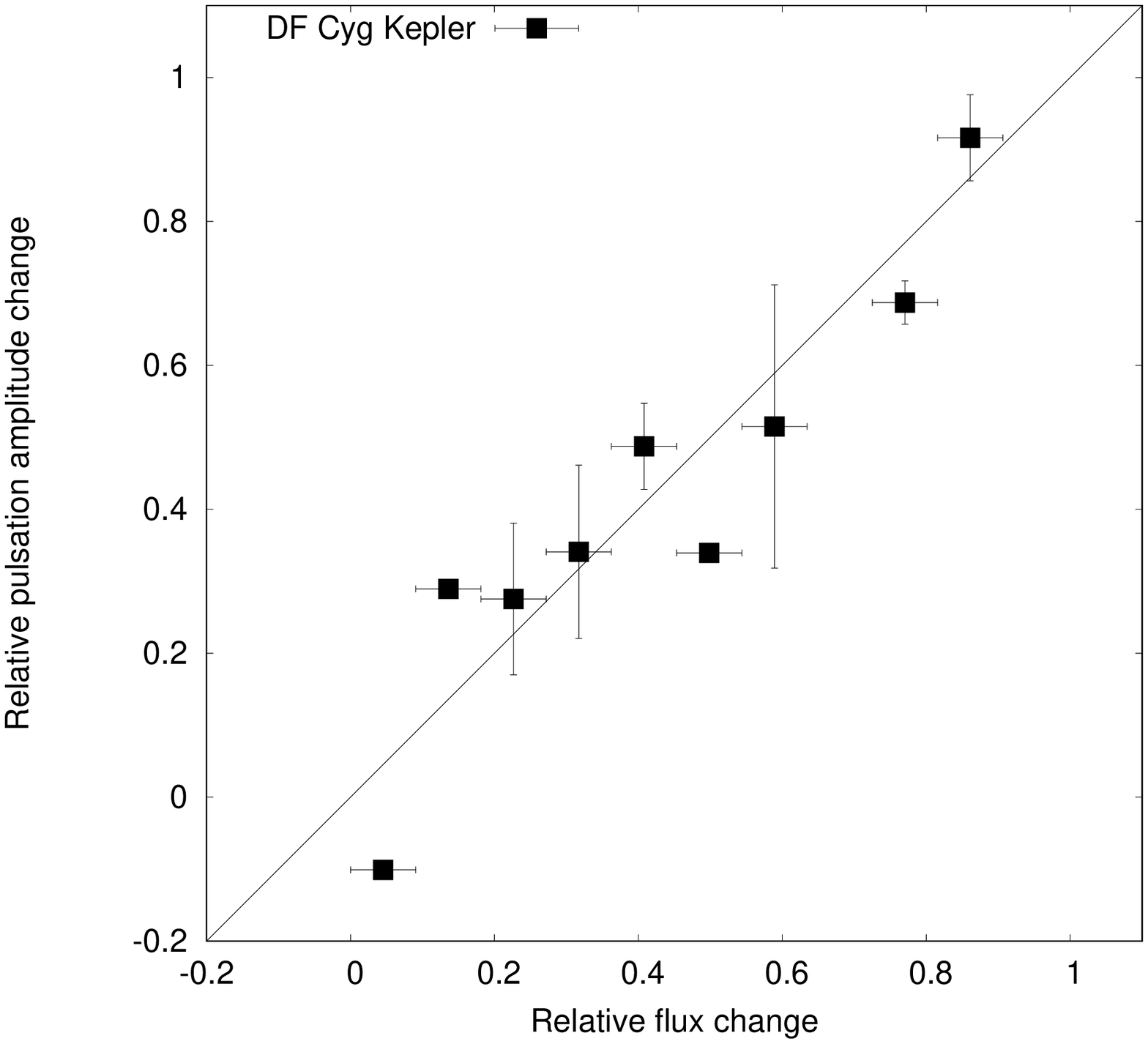}
    \caption{Pulsation amplitudes as function of the mean brightness for the {\it Kepler} data of DF~Cyg. {\it Top panel:} instantaneous flux amplitudes vs. mean fluxes in  linear scaling. The dotted red line shows a linear fit to the data, which was used to estimate the $f_\lambda^{\rm max}$ and $a_\lambda^{\rm max}$ parameters.  The formal error bars are smaller than the symbol sizes. {\it Bottom panel:} the relative pulsation amplitude vs. the relative mean flux in equidistant bins. Here the vertical error bars show the standard deviations of the mean, while the horizontal error bars indicate the width of the bins (note the missing bin at 0.7 due to lack of points). The diagonal black lines show the line of equality in both panels. See text for the details.} 
   \label{kepler-dfcyg}
   \end{figure}

We treated the four types of data (visual, OGLE $I$-band, ASAS $V$-band, { \it Kepler}) separately, partly because of the wavelength dependence of the pulsation amplitudes (Pollard et al. 1996), partly to avoid misleading effects from mixing heterogeneous data.

\begin{figure}
   \centering
   \includegraphics[width=9cm]{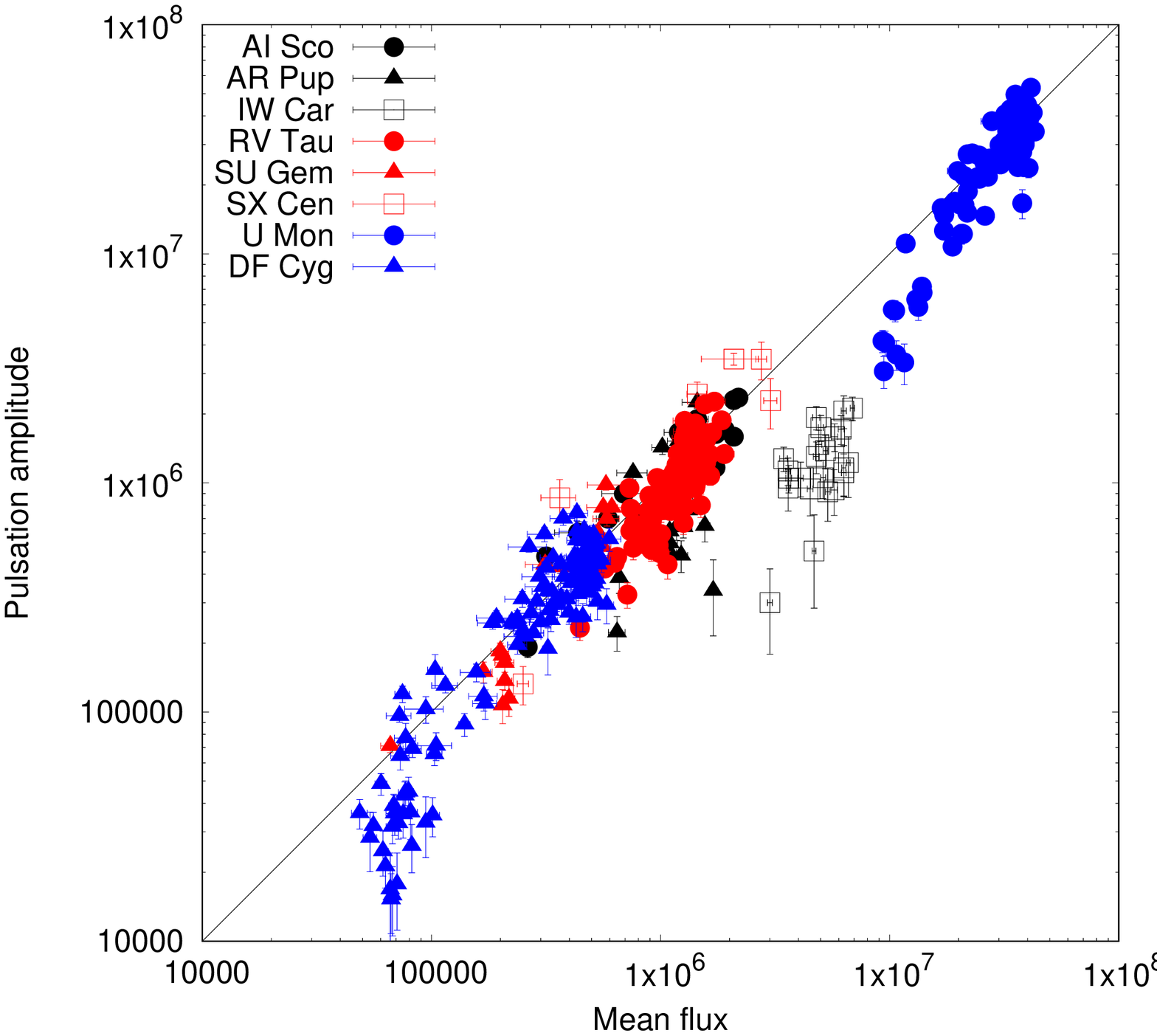}
   \includegraphics[width=9cm]{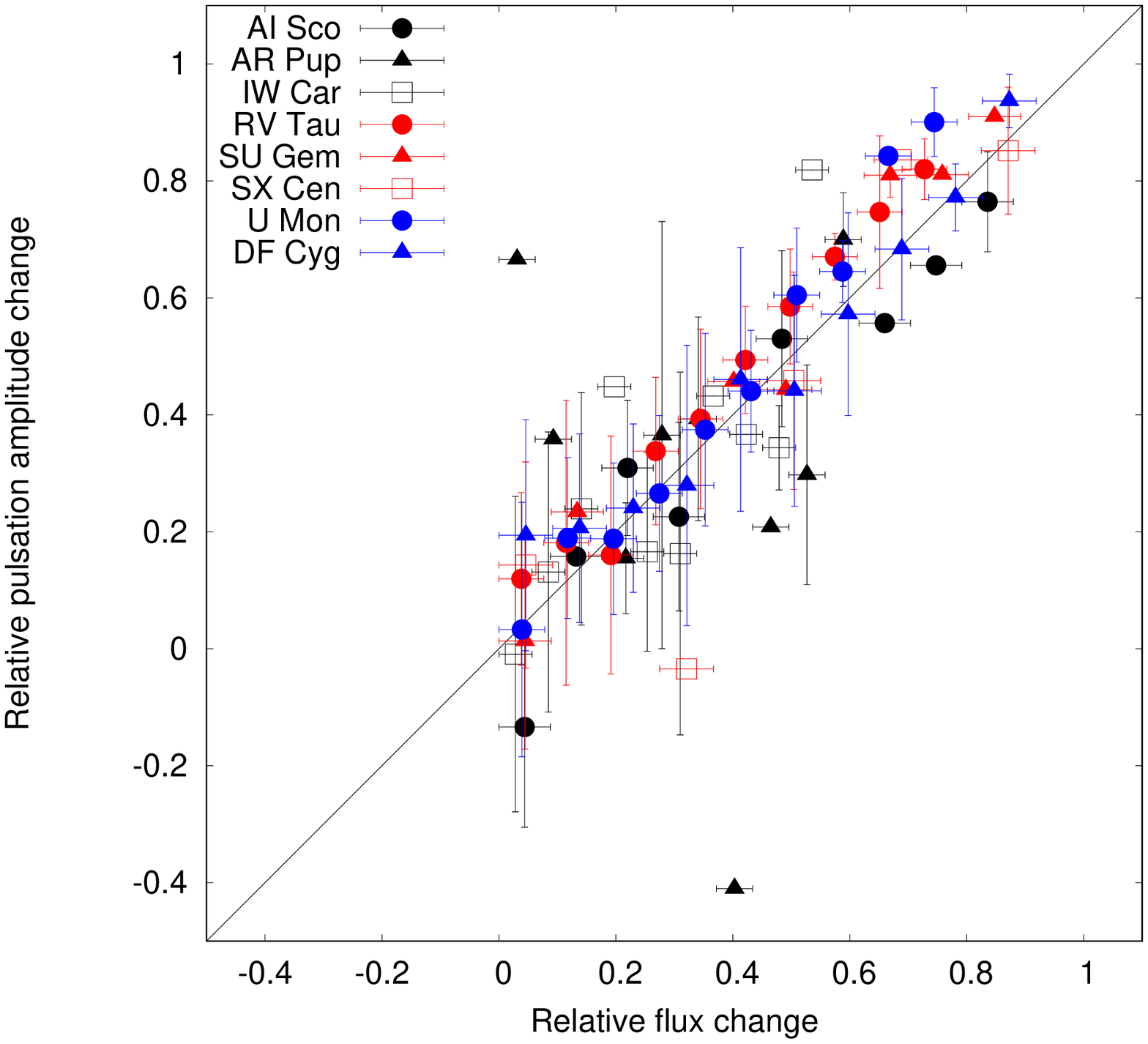}
   \caption{Pulsation amplitude as function of the mean brightness for the eight RVb stars from the AAVSO database (for each star, one point corresponds to one full pulsation cycle). Note the logarithmic scaling in the upper panel which allows easy comparison of the bright and the faint stars. The bottom panel is the same as in Fig.~\ref{kepler-dfcyg}.}
   \label{ampli-aavso}
   \end{figure}

\begin{figure}
   \centering
    \includegraphics[width=9cm]{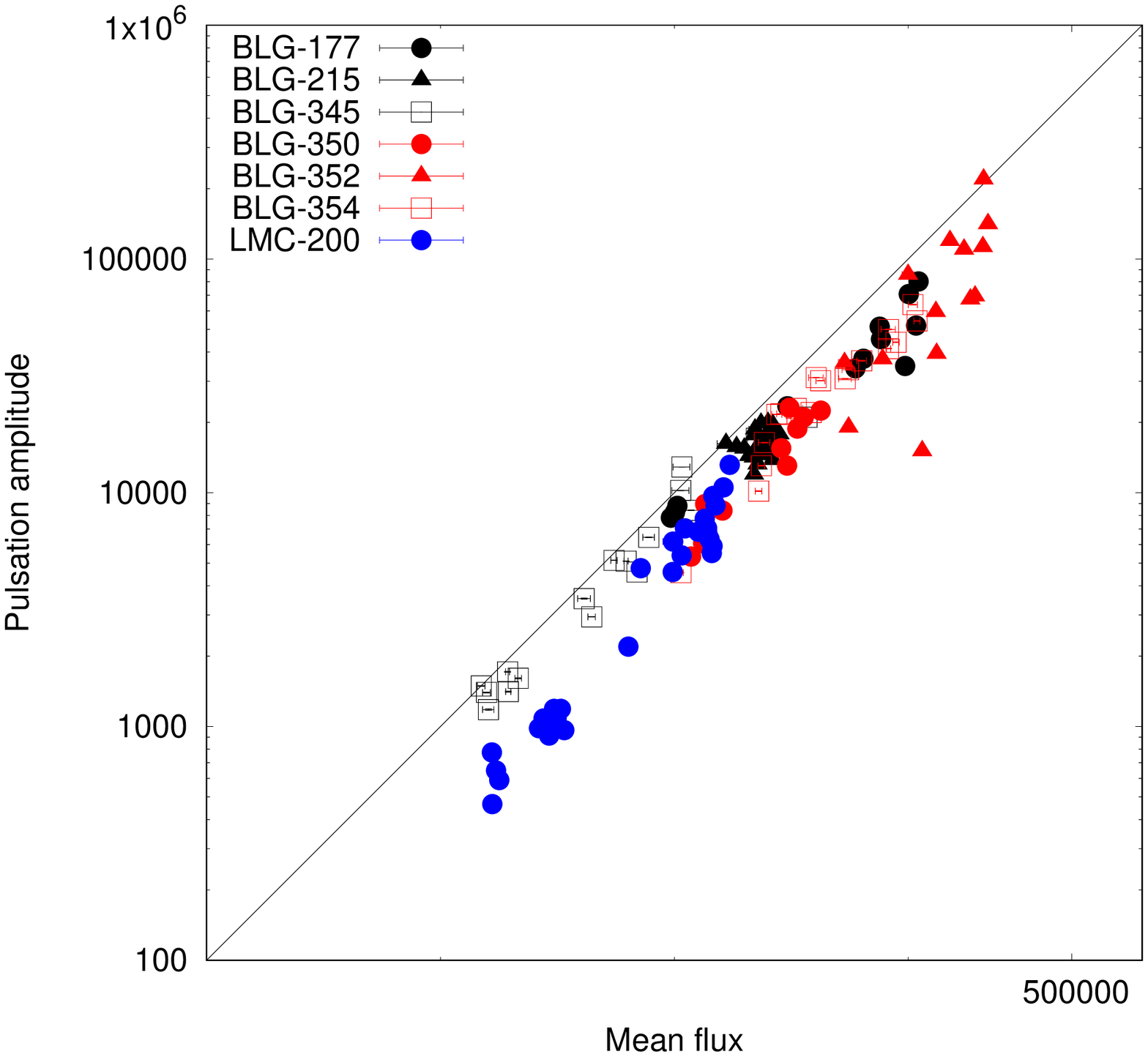}
    \includegraphics[width=9cm]{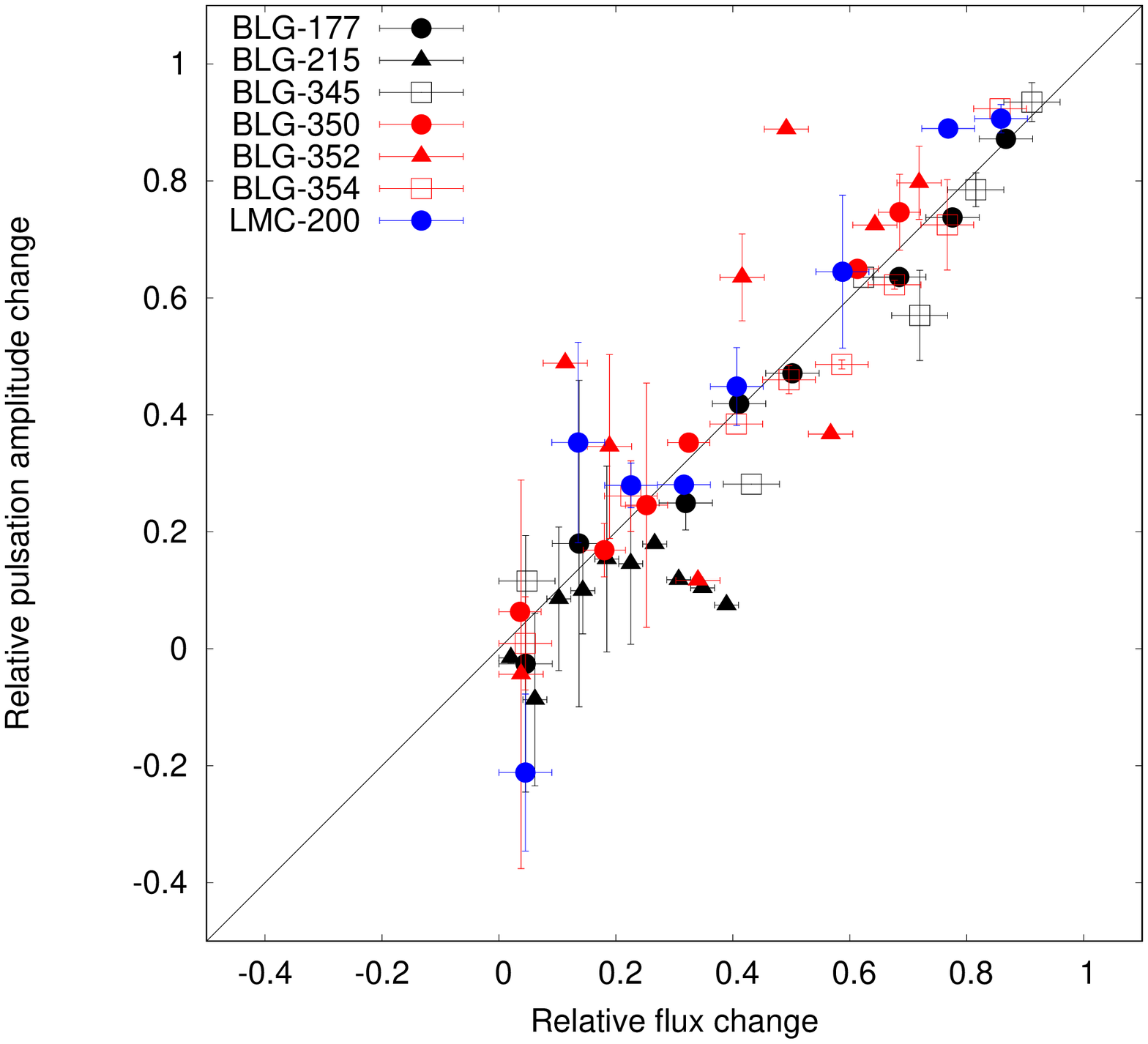}
    \caption{The same as in Fig.~\ref{ampli-aavso} for the seven OGLE stars.} 
   \label{ampli-ogle}
   \end{figure}

\begin{figure}
   \centering
   \includegraphics[width=9cm]{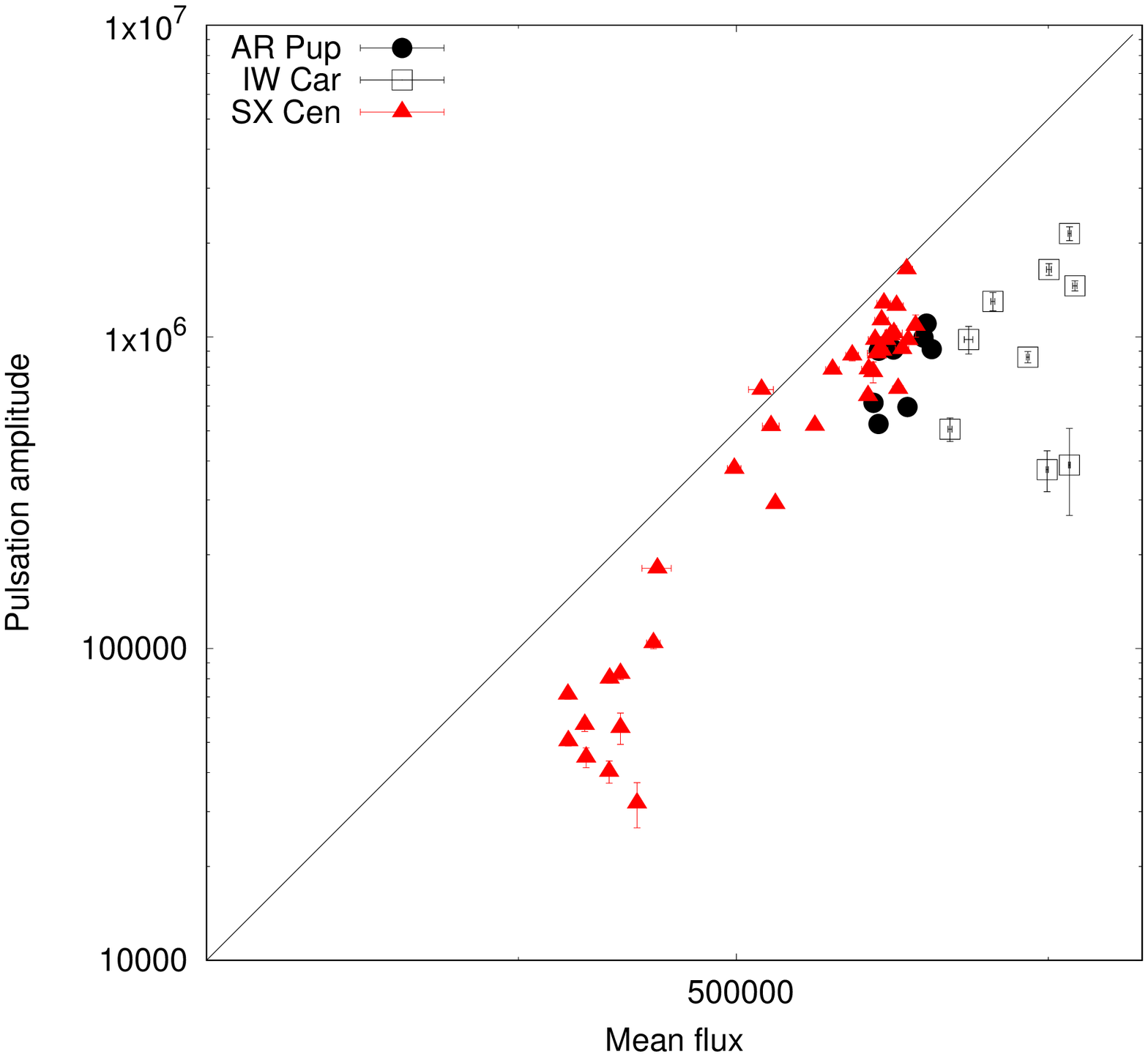}
\includegraphics[width=9cm]{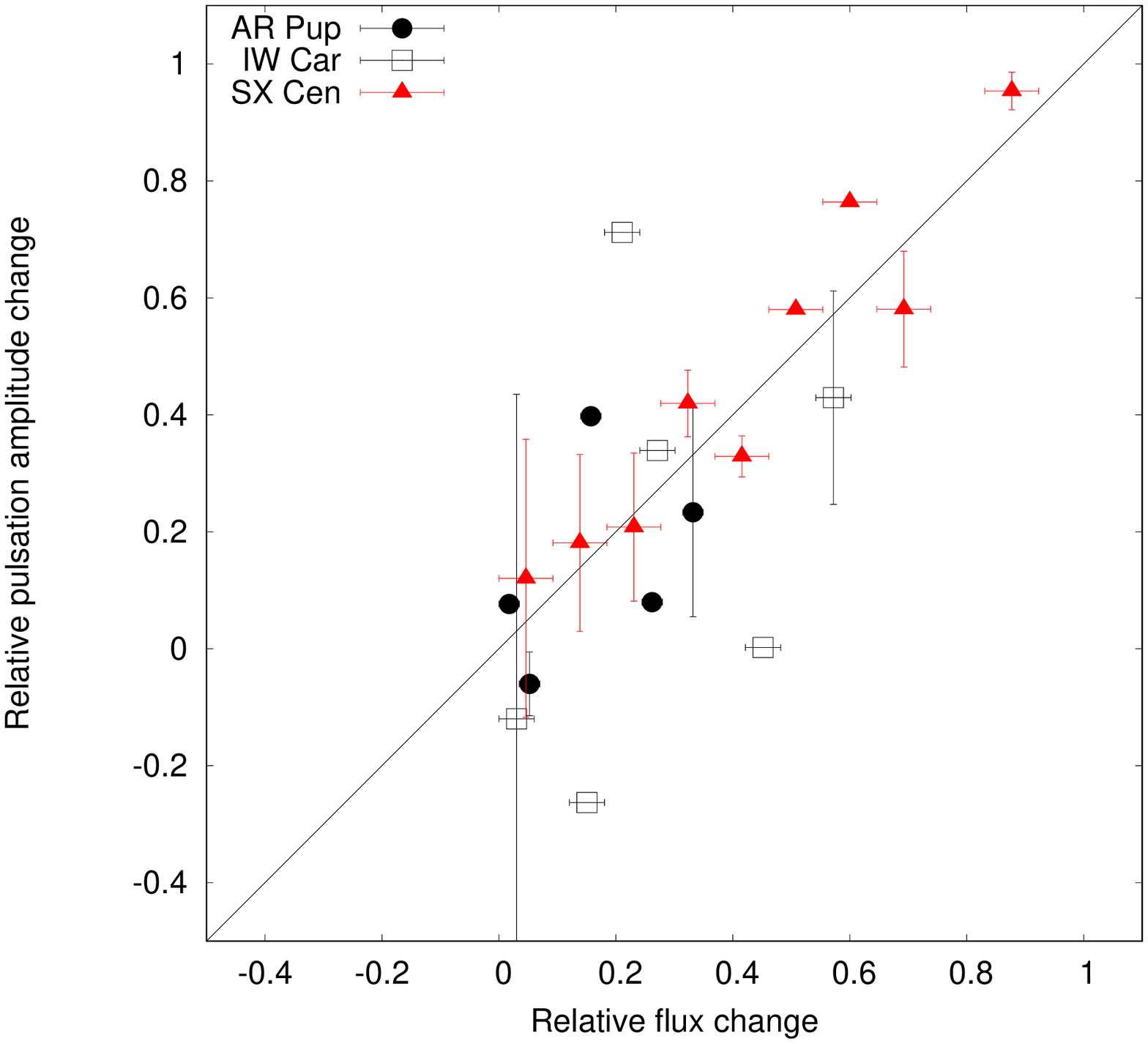}
   \caption{The same as in Fig.~\ref{ampli-aavso} for the three ASAS stars.}
   \label{ampli-asas}
   \end{figure}

First we show the raw correlations between the mean fluxes and the corresponding full amplitudes in the upper panels of Figs.~\ref{kepler-dfcyg}-\ref{ampli-asas}. For the sake of comparing bright and faint stars in Figs.~\ref{ampli-aavso}-\ref{ampli-asas}, 
we set doubly logarithmic scaling of the axes. To guide the eye, we have also drawn the lines of equality in these plots. While {\it Kepler} and visual data lie close to the equality, the OGLE and ASAS points all form parallel sequences falling below the diagonal lines, meaning that the correlations are close to linear but the mean slopes are less than one. The formal Pearson's $r$ correlation coefficients are greater than 0.8 for most of the stars except those three already mentioned in relation to Fig.~\ref{aavso} (IW~Car: $r= 0.53$ (AAVSO), AR~Pup: $r=0.21$ (AAVSO), OGLE-BLG-T2CEP-215: $r=0.34$).

The well expressed linear correlation between the amplitudes $a_\lambda$ and mean fluxes $f_\lambda$ for all types of the observations led us to fit the data in the upper panels of Figs.~\ref{kepler-dfcyg}-\ref{ampli-asas} with the simplest $a_\lambda=\alpha f_\lambda + \beta$ ($\lambda$=visual, $V$ and $I$) linear term. What we found is that the $\alpha$ slope parameters for almost all stars agreed within the error bars for each $\lambda$. Namely, the visual observations show a slope parameter around unity (e.g. AI~Sco: 0.95$\pm$0.10; RV~Tau: 0.89$\pm$0.08; SU~Gem: 1.21$\pm$0.14; U~Mon: 0.93$\pm$0.06), the OGLE $I$-band observations around 0.6 (e.g. BLG-177: 0.59$\pm$0.06; BLG-215: 0.68$\pm$0.09; LMC-200: 0.60$\pm$0.04), while the ASAS $V$-band data resulted in a meaningful correlation for SX~Cen only, with a slope of 0.62$\pm$0.08. The {\it Kepler} data of DF~Cygni show very similar linear amplitude vs. flux scaling throughout the whole RVb cycle than what the visual data imply. As we will demonstrate later, the wavelength dependence of the slope parameter follows that of the intrinsic pulsation amplitude (i.e. the smaller $I$-band slope parameter is a direct consequence of the pulsation amplitude being smaller in $I$ than in $V$).

In the next step we determined the relative flux change and the relative pulsation amplitude change as follows. For the fluxes, we defined their relative change as $(f_\lambda^{\rm max}-f_\lambda)/f_\lambda^{\rm max}$, where $f_\lambda^{\rm max}$ is the pulsation-averaged flux when the star is the brightest. This has defined the full range of relative flux changes, where 0.0 corresponds to the global maximum. To calculate the relative amplitude changes, we had to take into account the intrinsic cycle-to-cycle variability of the RV~Tau light curve shapes (well documented for RVa-type stars, e.g. R~Sct -- Buchler et al. 1996, AC~Her -- Koll\'ath et al. 1998). Because of that, it would have been misleading to take the single amplitude value that corresponds to the brightest flux point as the maximum amplitude; instead of that we fitted a line to the amplitude vs. flux plot and then took the fit's value for the largest flux as the hypothetic mean maximum amplitude $a_\lambda^{\rm max}$ (see the upper panel of Fig.~\ref{kepler-dfcyg} for a visualisation of these two parameters). Then the relative change of the amplitude was defined similarly to that of the flux, namely as $(a_\lambda^{\rm max}-a_\lambda)/a_\lambda^{\rm max}$. Finally, we calculated the mean relative amplitude change as a function of ten equally spaced flux bins that were selected between the minimum and maximum relative flux changes. 

The results are shown in the lower panels of Figs.~\ref{kepler-dfcyg}, \ref{ampli-aavso}, \ref{ampli-ogle} and \ref{ampli-asas} for the {\it Kepler}, AAVSO, OGLE and ASAS data, respectively. Here (0,0) corresponds to the brightest state with the largest amplitude. Note that the reason why generally there is no  point in the origin is twofold: (i) the bins are centered on their midpoints and (ii) the definition of the maximum amplitude implies individual points in the bright states to scatter around the origin. Two major conclusions can be drawn from these plots. First, within the error bars, the relative amplitude changes scale perfectly with the relative flux changes, which means that the pulsation amplitude, in fact, remains constant throughout the RVb cycle when compared to the overall system flux. The diagonal lines in the panels are not fits but indicate the lines of equality. All but a few RVb stars show the same proportionality than the one found for DF~Cygni by Vega et al. (2017) and this scaling is valid not only for the extrema (what actually was noticed by those authors) but it holds throughout the whole RVb cycle. Second, the scatter in the plots is rather dominated by the stars than by the observational uncertainties. This is illustrated by the {\it Kepler} plots in Fig.~\ref{kepler-dfcyg}, where the individual points in the upper panel have practically no measurement errors. It is the RV~Tau-type pulsation that changes seemingly irregularly (presumably due to strong non-linear effects, B\'odi et al. 2016) in the same range than the apparent scatter around the diagonal lines in Figs.~\ref{kepler-dfcyg}-\ref{ampli-asas}. What these mean is that the simplest explanation shall invoke a mechanism that equally affects the mean brightness and the apparent amplitude.
   
\section{Discussion}

Our results show a general pattern for the overwhelming majority of the sample, that is a linear correlation between the mean brightness and the pulsation amplitude, when everything measured in flux units. The one-to-one correspondence of the relative changes, depicted in the lower panels of Figs.~\ref{kepler-dfcyg}-\ref{ampli-asas}, is actually rephrasing the linear correlation differently, which can be understood as follows. Let us consider the equality of the relative changes of the pulsation amplitudes and the mean fluxes

\begin{equation}
\frac{\Delta a_\lambda}{a_\lambda}=\frac{\Delta f_\lambda}{f_\lambda}  
\end{equation}

\noindent as an approximation of the following differential equation

\begin{equation}
\frac{da_\lambda}{a_\lambda}=\frac{df_\lambda}{f_\lambda}. \end{equation}

\noindent This can be easily integrated as 

\begin{equation}
\log a_\lambda=\log f_\lambda + c_\lambda,
\end{equation}

\noindent where the integration constant $c_\lambda$ can be expressed as 

\begin{equation}
c_\lambda=\log \frac{a_\lambda^{\rm max}}{f_\lambda^{\rm max}}.
\end{equation}

\noindent Here $a_\lambda^{\rm max}$ and $f_\lambda^{\rm max}$ are the amplitude and the mean flux values when the star is the brightest. Substituting $c_\lambda$ into Eq.\ 4 and rearranging leads to the following formal solution:

\begin{equation}
a_\lambda=\frac{a_\lambda^{\rm max}}{f_\lambda^{\rm max}} f_\lambda,
\end{equation}

\noindent which is exactly the same linear relationship between the amplitudes and the fluxes as we have seen in the empirical data. The $\alpha$ slope parameter turns out to be the ratio of the maximal amplitudes and fluxes, which can be determined for each star and each photometric band separately. Given that the RV~Tau pulsations follow the same wavelength dependence in the photometric amplitudes as, for example, the Cepheid variables, the smaller $I$-band slope parameters found for the OGLE-stars are a direct consequence of having smaller pulsation amplitudes in $I$ than in $V$ (e.g. Pollard et al. 1996).   

Having established these very simple properties of the flux-amplitude relationship, one can ask about the implications. It has been noticed very early for individual stars that the pulsation amplitude shows some correlation with the long-term changes (for instance, O'Connell 1946 already noted for IW~Car that the pulsations have larger amplitudes when the star was brighter). A connection between the circumstellar material and the RVb phenomenon was proposed by Lloyd Evans (1985) who argued against the role of binarity, preferring unstable, potentially pulsation-induced mass-loss that can lead to R~Coronae~Borealis-like obscuration events. Fokin (1994) considered the problem of the secondary variability in RVb stars and, after discussing the difficulties of modelling it with some sort of pulsations, concluded that binarity shall play an important role instead, with quasi-periodic eclipses by a circumstellar dust cloud (similarly to the proposal of Waelkens \& Waters 1993). Zsoldos (1996) too argued against pulsational origin of the RVb phenomenon in RV~Tau but noted that simple binarity is also excluded because of the long-term changes of the RVb cycles.  Pollard et al. (1996), based on extensive long-term multicolour photometry of RV~Tau stars, noted various features of the RVb phenomenon. Two of their stars, U~Mon and AI~Sco, exhibited well-expressed `amplitude damping' during the RVb minima. While binarity was suggested for these stars, they also noted as an argument against dust obscuration that ``pure obscuration by dust will give a reddening and a dimming of the obscured star but should not decrease the amplitude of the pulsations or make the deep-shallow alternations less distinct''. This line of argumentation was adopted by later authors, like Van Winckel et al. (1999) and  Maas et al. (2002). 

We think it is clear that the recurrent reasoning in the literature against dust obscuration is not correct. It was only very recently, that Vega et al. (2017) pointed out that this kind of amplitude damping is actually what one expects for the pulsating stellar disk being occulted during the long-period minima by a very large, opaque screen, which, they argue, should be a circumbinary dusty disk around the entire binary system. Our results presented in this paper indicate that amplitude variability is ubiquituous and follows the same linear scaling with the mean flux in almost every RVb star. In other words, the pulsation amplitude remains constant (with some intrinsic small-scale variability due to the non-linear nature of the pulsations), when measured relative to the overall system brightness. One of the reasons why this has not yet been found is the general use of the magnitude system in variable star analyses: the inverse logarithmic nature hid the simple relationship between the mean brightness and the pulsation amplitude. Also, the fluctuating amplitude variability of RV Tau-type stars did not help either: one needs very good duty cycle and long time-span to average out the effects of the cycle-to-cycle changes that are inherent to the RV~Tau-type pulsations.  

Finally, we briefly turn to those stars that have marginally similar behaviour than the whole sample. These are IW~Car, AR~Pup and OGLE-BLG-TCEP2-215. The classification of RV~Tau stars have long been known to be very difficult (e.g. Zsoldos 1998), with semiregular (SR) variables frequently mistaken for other types of luminous variables. The periods, amplitudes and systematic amplitude changes of SR variables (Kiss et al. 2000) can indeed by similar photometrically, hence further information is always important. For IW~Car, a rotating and expanding post-AGB nebula was recently resolved by ALMA (Bujarrabal et al. 2017), which is the latest update to its long known post-AGB status. However, its missing light curve alternation means that the star is more like a general pulsating post-AGB star than a classical RV~Tau-like variable. This is in line with the fact that Giridhar et al. (1994) derived a spectroscopic effective temperature of 6700 K, which is much hotter than typical RV~Tau stars (Kiss et al. 2007). AR~Pup and its post-AGB disk was observed interferometrically by Hillen et al. (2017), who noted that for this star, the total infrared luminosity dominates over the dereddened optical fluxes, which is indicative of the disk close to edge-on. The least is known for the OGLE star, for which only the very red colour ($V-I\approx3.3$ mag) is listed in the catalogues, which could indicate both high interstellar reddening and instrinsically red colour. We speculate that these three stars may have different geometry and/or circumstellar extinction than the rest of the sample. Even for IW~Car and AR~Pup, there are cycles of the RVb variability when there are hints of the amplitudes following the variations of the mean flux levels. The RVb cycles are not strictly repetitive in other stars either (Zsoldos 1996), implying that the obscuring clouds are changing over the time-scale of the orbital periods of the systems.  All in all, while these stars exhibit a somewhat noisier relationship than the other stars, fundamentally they still exhibit a similarly linear relation between pulsation amplitude and system flux, and therefore fit the proposed general picture. 

\section{Summary}

The main results of the paper can be summarized as follows:

\begin{enumerate}
    \item We have compiled light curves for a sample of RVb-type variables, using visual observations, ground-based CCD photometric measurements and ultra-precise data from the {\it Kepler} space telescope.
    
    \item We found a ubiquitous linear correlation between the pulsation amplitude and the mean brightess, when both are measured in flux units. There is a one to one correspondance between their relative changes, meaning that the pulsation amplitude actually remains constant throughout the RVb cycle, when measured relative to the system flux level.
    
    \item The properties of the correlation can be naturally explained by a mechanism that equally affects the mean flux and the  apparent amplitude, so that the whole light curve is scaled by a time-dependent factor. Periodically variable obscuration by a large opaque screen, presumably corresponding to a circumbinary dust disk, provides the required mechanism. 
    
    \item We conclude that the light variations of RVb-type stars can be fully explained phenomenologically by the combination of time-dependent non-linear pulsations and the dust obscuration model of the RVb phenomenon. 
\end{enumerate}

\begin{acknowledgements}
This work has been supported by the NKFIH K-115709 and the GINOP-2.3.2-15-2016-00003 grants of the Hungarian National Research, Development and Innovation Office, and the Hungarian Academy of Sciences. This research has made use of the International Variable Star Index (VSX) database, operated at AAVSO, Cambridge, Massachusetts, USA. We acknowledge with thanks the variable star observations from the AAVSO International Database contributed by observers worldwide and used in this research. This paper includes data collected by the Kepler mission. Funding for the Kepler mission is provided by the NASA Science Mission directorate. We thank an anonymous referee for his/her useful comments and suggestions.
\end{acknowledgements}

\end{document}